\documentclass[12pt]{article} 

\usepackage{amsfonts}
\usepackage{amsmath}
\usepackage{amsthm}
\usepackage{amssymb}
\usepackage{color}
\usepackage{graphicx}
\usepackage{enumerate}
\usepackage{enumitem}

\usepackage[margin=1in]{geometry}

\def\btheta{{\mathbf{\theta}}}
\def\balpha{{\mathbf{\alpha}}}
\def\bbeta{{\mathbf{\beta}}}
\def\calR{{\mathbb{R}}}
\def\bX{{\mathbf{X}}}
\def\bH{{\mathbf{H}}}
\def\bx{{\mathbf{x}}}

\def\bb{{\mathbf{b}}}

\def\argmin{\mathop{\rm argmin}}

\begin{document}

\title{Robust nonparametric regression: review and 
practical considerations.}

\author{Matias Salibian-Barrera} 

\date{March 17, 2023}

\maketitle

\abstract{Nonparametric regression models offer a way to 
understand and quantify relationships between variables without 
having to identify an appropriate family of possible 
regression functions. Although many estimation methods for these models 
have been proposed in the literature, most of them can be highly sensitive
to the presence of a small proportion of atypical observations in the 
training set.  A review of outlier robust estimation 
methods for nonparametric regression models is provided, 
paying particular attention to practical considerations. 
Since outliers can also influence negatively the regression estimator by 
affecting the selection of 
bandwidths or smoothing parameters, a discussion of robust
alternatives for this task is also included. 
Using many of the ``classical'' 
nonparametric regression estimators (and their robust counterparts)
can be very challenging in settings with a moderate or large number of
explanatory variables, so recently proposed 
robust nonparametric regression methods that scale well with a growing number
of covariates are also discussed.}

\section{Introduction}

Regression models can be used to either understand and quantify the relationship between a 
certain variable 
$Y \in \mathbb{R}$ (generally called the ``response'') 
and a number of ``explanatory" ones (or ``features'')
$\bX$, 
or to construct predictions for $Y$ 
based on future values of $\bX$. Note that typically
$\bX \in \mathbb{R}^d$, $d \in \mathbb{N}$, but  
some models and estimation methods 
can be extended naturally to more general cases where 
$\bX \in {\cal X}$, a metric space (e.g. $\bX$ could be
an image or a smooth function). 
In what follows we will consider a general regression model 
of the form
\begin{equation} \label{eq:model1}
Y = f(\bX) + \varepsilon \, ,
\end{equation}
where $f$ is a real valued unknown function, 
and $\varepsilon$ is a random variable that 
models random fluctuations not  
accounted by $f$. 
Furthermore, we will be concerned with
the possible presence of atypical observations 
or with heavy-tailed
error distributions. Thus, we 
will avoid requiring that moments of $\varepsilon$ necessarily exist, 
and instead assume that the conditional 
distribution function of $\varepsilon$ given $\bX$ is of the form
\begin{equation} \label{eq:cond.disp}
F_{\varepsilon | \bX}(u) = F_0( u / \sigma_{\bX} ) \, ,
\end{equation}
for a fixed distribution function $F_0$ that is symmetric 
around zero, and where 
$\sigma_{\bX} > 0$ is the error scale parameter, which is
typically unknown. 
Note that \eqref{eq:cond.disp} is a particular case of a more
general formulation that would specify 
$F_{\varepsilon | \bX}(u) = F_0( u , \bX )$. 
The setting on which we focus implies that any heteroskedasticity 
in the data is due to a changing dispersion coefficient (e.g. standard 
deviation, if second moments exist), rather than more general changes in 
the shape or family of the error distribution, for example. Although 
the particular functional form of the dependence of the distribution 
of $\varepsilon$ conditional on $\bX$ is not critical, 
we will see below
that to be able to use robust estimators for these models, the
mechanism of this dependence will need to be assumed 
explicitly, in order for the residuals in the estimating equations to be 
adjusted appropriately.


A widely used class of regression models assumes that
the regression function $f$ in \eqref{eq:model1} belongs to 
a family parametrized by  a finite dimensional vector: 
${\cal F} = \left\{ f_{\mathbf \theta}, \, \btheta \in \Theta \subseteq
\mathbb{R}^k \right\}$. In this case, the problem
of estimating $f$ reduces to that of estimating  the vector $\btheta$. 
A clear limitation of adopting this 
framework is that, to avoid drawing biased conclusions
due to model misspecification, the family ${\cal F}$ needs to 
be rich enough to include the unknown 
$f$ in \eqref{eq:model1}. Although in some applications 
subject matter knowledge
(e.g. physical models) may prescribe an appropriate family
${\cal F}$, there are many practical situations where
no obvious parametric model exists and 
a different approach is necessary. 

Nonparametric regression analysis refers to the collection of estimation  
methods for $f$ that do not assume a finite dimensional
parametric model for it. 
Nonparametric methods can coarsely be classified 
into two groups, which we will call ``unstructured" and 
``semi-structured", respectively. 
 ``Unstructured" approaches place almost no 
asumption or restriction on the shape or structure of the regression
function or its estimator. An example is given by local (kernel) 
regression methods.  ``Semi-structured" approaches, on the other hand, 
impose some structure on $f$, or, more often, on its estimator $\hat{f}$.  
Examples in this latter class include additive models, 
and regression trees. These constraints
generally result in estimators that may be notably 
less variable than ``unstructured'' ones, and thus 
have lower prediction errors if the bias 
induced by the added structure is sufficiently small. 
This coarse taxonomy of non-parametric regression
estimators is not meant to be exhaustive, and 
there are methods and models that do not fall clearly in either
of these two groups. However, we believe that it is 
useful to organize our discussion below. 

There are also some models that are neither fully parametric nor
completely nonparametric. These are often called semiparametric in the 
literature. One popular class of such models are
index models, where $\bX \in \mathbb{R}^d$ with $d > 1$, 
and the regression function is assumed to satisfy
$$
f( \bX ) = \mu + \sum_{j=1}^p f_j( \bX^\top \balpha_j ) \, ,
$$
where $\mu \in \calR$, $f_j : \calR \to \calR$ are smooth functions 
of a single argument, and 
$\alpha_j \in \calR^d$. Obvious ambiguity in the expression above can be 
avoided if one assumes that $f_j(0) = 0$, $\| \balpha_j \|_2 = 1$, 
and that the first non-zero element of $\balpha_j$ is positive, for all
$1 \le j \le p$. However, identifiability of these models is a delicate problem 
(Horowitz, 1998, page 14). We refer the interested
reader to Yuan (2011) and references therein. 
Robust estimators for these models proposed in the literature focus on estimating the 
projections $\balpha_j$. 
Another class of semiparametric models is that of partly linear models where
$f$ is assumed to depend linearly on some covariates and also 
include a non-parametric
component: 
$$
f( \bX, u ) = \bbeta^\top \bX + g( u ) \, ,
$$
where $\bbeta \in \calR^d$ and $g : \calR \to \calR$
is a smooth function.  Robust estimators for these models are typically
obtained by combining robust linear regression estimators with some 
of the nonparametric approaches discussed below. 

%
%
%
%

Historically, the focus in the Statistics literature 
when it comes to regression problems
has been on estimating the function $f$. 
Different approaches are generally compared with each other based on 
measures of uncertainty
associated with the estimated values of $f$ at specific 
points of its domain, for example the mean 
squared error at $\bx_0$:
\begin{equation} \label{eq:mspex0}
E \left[ ( f(\bx_0) - \hat{f}(\bx_0) )^2 \right] \, ,
\end{equation}
or with overal error measures such as the
integrated squared error (ISE) 
\begin{equation} \label{eq:ise}
\mbox{ISE} = 
\int ( f(\bx ) - \hat{f}(\bx) )^2  dF_\bX(\bx)  \, ,
\end{equation}
where $F_\bX$ denotes the distribution function of the explanatory
variable $\bX$. 
In recent decades many methods 
have been proposed to construct ``predictors'' for
the response variable $Y$ at given values of $\bX$ while remaining relatively agnostic 
about the regression function $f$ (e.g. random forests and 
gradient boosting). 
The main difference between estimation-based and prediction-based
approaches lies in 
how one quantifies the ``performance" of a specific method. 
Nevertheless, there is of course a natural connection between 
these two approaches: given an 
estimator $\hat{f}$ and a point $\bx_0$, a natural
prediction for $Y$ 
when $\bX = \bx_0$ 
is $\hat{f}(\bx_0)$. 
Similarly, one can estimate $f(\bx_0)$
using a predictor 
by calculating the prediction $\hat{Y}$ for $\bX = \bx_0$. 

Estimators for the regression function $f$ 
are constructed using what is often called a ``training
set''. Namely, we assume that we have access to a 
data set ${\cal D}_n = \left\{ (y_i, \bx_i), 1 \le i \le n \right\}$
that follows the model in \eqref{eq:model1}. 
This assumption can be stated formally by considering 
a random object $(Y, \bX)$ where 
$Y = f(\bX) + \varepsilon$ as in \eqref{eq:model1}
and assuming 
that the points in ${\cal D}_n$ are a random sample from 
$(Y, \bX)$. 
The process
of constructing $\hat{f}$ (an estimator, or a predictor) using ${\cal D}_n$ is called 
``estimation'' in the traditional Statistics literature, and ``learning''
in the Machine Learning literature. In what follows we will use these
terms exchangeably. 

In the Statistics literature, nonparametric methods traditionally have 
been called ``robust'' 
in the sense of requiring fewer or weaker assumptions than their 
parametric counterparts to be consistent to the parameter or
function of interest. 
However, these methods typically still require that the 
points in the training data are 
instances of the same random object (i.e. that every observation in 
${\cal D}_n$ comes from the same population). 
In this paper we are concerned with methods that remain informative
even when a proportion of the training sample may follow a
different distribution.  We will say that a method is 
``robust''  when it is not unduly affected by a relatively 
small proportion of atypical observations in ${\cal D}_n$. It is well 
known that many non-parametric regression estimators and predictors 
are not robust in the latter sense. 
In other words, a few points in ${\cal D}_n$ 
that deviate from \eqref{eq:model1}
may have a very large (even unbounded) effect on the estimator $\hat{f}$ or 
predictions $\hat{f}(\bx_0)$. Robust methods are designed to avoid this 
problem while also maintaining a good performance when there are no ``outliers'' 
in the training set. 
The purpose of this paper is to review available 
robust alternatives for different non-parametric regression estimators.  

We focus here on estimation and prediction methods to
estimate the regression function $f$ as in \eqref{eq:model1} when the 
errors $\varepsilon$ have a symmetric distribution around zero. 
In this case, $f(\bx_0)$ is the ``center'' of the conditional distribution  
of $Y$ given $\bX  = \bx_0$. 
In the interest of space, regression analysis methods with a 
different goal (e.g. quantile regression or modal
regression) will not be included in this review. 

Note that the infinite dimensional nature of the ``parameter'' 
space in nonparametric regression models limits the 
direct application of many of the
standard robustness measures in the literature (e.g. breakdown point). 
There are few attempts in the literature
to formalize and quantify the notion of robustness for these models.  
Tamine (2002) approached the problem in a point-wise manner, and 
derived smoothed influence functions for 
Nadaraya-Watson type estimators of $f(x)$ at each fixed point $x$. 
Giloni and Simonoff (2005) studied point-wise breakdown points of
$L_1$ local polynomial regression estimators. 
Christmann and Steinwart (2007)
and Steinwart and Christmann (2008, Section 10.4) 
studied the influence function of the \textit{function estimate}
$\hat{f}$ obtained using
kernel-based regression (i.e. penalized regression over a 
reproducing kernel Hilbert space (RKHS) of functions with a bounded 
and continuous kernel). They found sufficient conditions for
these influence functions to be bounded. 
Hable and Christmann (2011) and Christmann et al. (2013)
showed that these RKHS estimators are also ``stable'' 
in the sense of being qualitatively robust
(Hampel, 1971). 
Note that there is a natural ``tension'' between 
stability (qualitative robustness) and consistency, 
which intutitively is related to the richness of the 
``parameter space'' (Hable and Christmann, 2013). 

The rest of the paper is organized as follows. 
Sections 2 and 3 discuss ``unstructured'' and ``semi structured''
methods, respectively, along with their robust alternatives. 
Some reflections on the conceptual challenges of 
studying robustness in a non-parametric setting
are included in Section 4.  

\section{Unstructured methods} \label{sec:unstr}

Kernel-based estimation methods impose very weak 
structures on the estimated $f$'s. 
For example, although a local constant (Nadaraya-Watson) kernel estimator
$\hat{f}(x_0)$ is a linear function of the observed responses in the
training set, the coefficients of the linear
combination change non-linearly as functions of 
$x_0$, so the shape of $\hat{f}$ as a whole cannot be easily
characterized. Penalized spline regression 
estimators (and regularized sieves (Grenander, 1981) estimators in general)
are somewhat restricted (since $\hat{f}$ is a linear combination
of the elements of the basis), but the spanned subspaces are
usually quite complex and the dimensions of the bases increase with
the training sample size, and thus it is not generally possible to 
describe explicitly the resulting family of 
estimators $\hat{f}$'s. 

In this section we 
discuss both kernel and spline robust estimators for
the regression function $f$ in \eqref{eq:model1}, 
focusing on the case where both $Y$ and $\bX$ are 
random objects. Robust fits based on 
local regression and splines methods 
can intuitively be constructed by simply
replacing the squared loss function with one that either
grows more slowly 
or is bounded. We show below that these arguably 
``ad hoc'' approaches
can in fact be justified formally from first principles.

\subsection{Kernel-based methods}

Under well known and relatively weak regularity
conditions 
the regression function 
$f$ is the conditional expectation 
of $Y$ given $\bX = \bx$: 
$$
f(\bx) = E \left( Y | \bX = \bx \right) \, .
$$
This observation suggests that, given a sample ${\cal D}_n$, 
a ``natural'' estimator for $f(\bx_0)$ 
is the Nadaraya-Watson local mean kernel 
estimator (Nadaraya, 1964; Watson, 1964). 
To fix ideas, consider first the case where $\bX \in \mathbb{R}$, then 
\begin{equation} \label{eq:nadaraya}
\hat{f}(x_0) = \frac{ \sum_{i=1}^n K_h( x_i - x_0 ) \, y_i}{\sum_{j=1}^n
K_h( x_j - x_0 ) } \, , 
\end{equation}
where the weigths $K_h( x_i - x_0) = (1/h) K( (x_i- x_0)/h)$ depend on 
a kernel function $K$, which is a
non-negative and symmetric function that satisfies 
$\int K(u) du = 1$, and the bandwidth $h > 0$. 
For example, the Epanechnikov kernel is given by 
$$
K(u) = \frac{3}{4} \left( 1 - u^2 \right) I( |u| \le 1) \, ,
$$
where $I(A)$ is the indicator function of the set $A$. 
See H\"ardle et al. (2004)
for more details. 
Note that $\hat{f}(x_0)$ above satisfies
\begin{equation} \label{eq:local.const}
\hat{f}(x_0) = \argmin_{a \in \calR} \ \sum_{i=1}^n
K_h \left(  x_i - x_0 \right) \,   
\left( Y_i - a \right)^2 \, . 
\end{equation}

These 
estimators can be extended rather naturally to the case where
$\bX \in \mathbb{R}^d$ by using a multivariate kernel
function $K_H$:
\begin{equation} \label{eq:multiv.local.const}
\hat{f}(\bx_0) = \argmin_{a \in \calR} \ \sum_{i=1}^n
K_{\bH} \left( \bx_i - \bx_0  \right) \,   
\left( Y_i - a \right)^2 \, . 
\end{equation}
where often the matrix $\bH = \text{diag}(h_1, h_2, \ldots, h_d)$, 
with individual bandwidths, $h_j > 0$,  
$K_{\bH}(\bb) = {\cal K}_d(\bH^{-1} \bb)$, and the 
multivariate kernel ${\cal K}_d$ is the product of univariate kernels: 
${\cal K}_d(\bb) = \prod_{j=1}^d K(b_j)$,  
for $\bb = (b_1, b_2, \ldots, b_d)^\top$, 
and $K$ is a univariate kernel as before. 
When the explanatory variables $\bX$ are only assumed to 
take values in an arbitrary metric space ${\cal X}$ 
with distance $m(\cdot, \cdot)$, then the above estimator
can be adapted in a natural manner, for example:
$$
\hat{f}(\bx_0) = \argmin_{a \in \calR} \ \sum_{i=1}^n
K_h \left( m(\bx_i, \bx_0)  \right) \,   
\left( Y_i - a \right)^2 \, . 
$$
However, the way in which the bandwidth parameter enters in the 
expression above 
will generally depend on the form of the metric $m$
(see, for example, Ferraty and Vieu, 2006). 

Formulations \eqref{eq:local.const} and \eqref{eq:multiv.local.const}
also suggest variants of these estimators using local polynomial
approximations. Indeed, when $\bX \in \calR^d$
better statistical properties (in terms of asymptotic bias)
can be obtained by using local linear approximations 
(see, for example, H\"ardle et al., 2004). Let 
\begin{equation} \label{eq:multiv.local.linear}
(\hat{\alpha}_{\bx_0}, \hat{\bbeta}_{\bx_0}^\top)^\top = \argmin_{ (\alpha, \bbeta) \in \calR^{d+1}} \ \sum_{i=1}^n
K_{\bH} \left( \bx_i - \bx_0  \right) \,   
\left( Y_i - \alpha - \bbeta^\top  ( \bx_i - \bx_0) \right)^2 \, , 
\end{equation}
and set $\hat{f}(\bx_0) = \hat{\alpha}_{\bx_0}$, where the kernel $K_{\bH}$ is as above. 
A Taylor expansion of the regression function plus 
these higher order local polynomial approximations can be used to obtain estimators of the 
derivatives of $f$. 

The bandwidth parameters play a critical role in determining the 
properties of the estimator $\hat{f}$. 
On the one hand, smaller bandwidths tend to produce ``wiggly'' estimators
that adapt closely to the points in ${\cal D}_n$
and may fail to generalize well to future observations,
resulting in both a poor estimate of $f$ and highly variable predictions. 
On the other hand, 
large bandwidths typically result in overly smooth $\hat{f}$'s that 
fail to reflect adequately 
the shape of the true regression function $f$, resulting in 
biased fits and predictions. 

Given a training set,
choosing an appropriate bandwidth is generally a challenging problem. 
Commonly used strategies include: (a) selecting $h = h_n$ 
in order to minimize either the leading term of the asymptotic mean squared error of $\hat{f}(x_0)$ in 
\eqref{eq:mspex0}, or the ISE in \eqref{eq:ise}; and (b) 
minimizing a direct estimate of the mean squared prediction error 
(generally via some variant of 
cross-validation). 
The former 
bandwidth choice is often called ``plug in optimal'', 
and it typically requires  
the estimation of several 
unknown quantities in the asymptotic mean squared error of $\hat{f}(x_0)$, 
such as the scale $\sigma_{x_0}$ of the 
conditional distribution of the errors, 
and the derivatives of the regression function $f$. 

Alternatively, one can select the value of $h > 0$ that minimizes 
an estimate of the prediction error. Cross-validation (CV) is 
a popular and practical alternative. 
Leave-one-out CV selects the value of $h > 0$ that minimizes 
\begin{equation} \label{eq:cv}
\mbox{CV}(h) = \frac{1}{n} \sum_{i=1}^n \left( y_i - \hat{f}_h^{(-i)}(x_i) \right)^2 \, ,
\end{equation}
where $\hat{f}_h^{(-i)}(x_i)$ denotes the estimate for $f(x_i)$ obtained with bandwidth $h$ 
and without using the point $(y_i, x_i)$ in ${\cal D}_n$. 
A variant of this approach with better statistical properties when applied to
local-linear estimators 
is the one-sided
cross validation (Hart and Yi, 1998). 

Note that 
$\mbox{CV}(h)$ is not an estimator
of ISE \eqref{eq:ise}, but rather of the expected squared prediction error 
\begin{equation} \label{eq:mspe}
	\mbox{MSPE} = E \left[ \left( Y - 
	\hat{f}(\bX) \right)^2 \right]  = E \left[ 
	E \left\{ \left. \left( Y - 
	\hat{f}(\bX) \right)^2 \right| {\cal D}_n  \right\} \right]
	\, ,
\end{equation}
where the expected value is taken over 
the random pair $(Y, \bX)$ and also the randomness 
in $\hat{f}$ resulting from the sampling variability in the training set 
${\cal D}_n$ 
(Hastie et al., 2009). Standard arguments show that if
second moments exist and the errors $\varepsilon$ in \eqref{eq:model1} 
satisfy $E[ \varepsilon | \bX = \bx ] = 0$ (for almost all $\bx$),
then minimizing \eqref{eq:mspe} is equivalent to minimizing the mean ISE
$$
\text{MISE} = E \left[ \int ( f(\bx ) - \hat{f}(\bx) )^2  dF_\bX(\bx)  
\right] = E \left[ \text{ISE} \right] \, ,
$$
where the expected value is taken over all possible training
sets. 
K-fold cross validation is similar to the above, except that the 
predictions are obtained by removing a randomly chosen block of  
observations from ${\cal D}_n$. Specifically, one randomly splits 
the training set into approximately equal-sized K blocks, 
and obtains predictions for the observations in each block using 
the data in all the other blocks. 
K-fold CV may be preferred over leave-one-out when 
predictions for observations in a block can be obtained relatively cheaply 
once a 
model fit has been computed using the other $K-1$ blocks of the training set 
(think linear regression models, for example). 
In this case, one essentially needs to ``fit the model'' (run the ``learning
algorithm'') only $K$ times to compute \eqref{eq:cv}, 
as opposed to $n$ times if using leave-one-out. 
However, for kernel local estimators, K-fold CV 
will typically have the same computational cost as its
leave-one-one version, since
\eqref{eq:local.const} or \eqref{eq:multiv.local.const} 
need to be solved
$n$ times regardless of the number of folds one uses.  


\subsubsection{Robust kernel / local regression estimators}

It is easy to see that a small proportion of outliers in the training set can have a
substantial direct negative effect on the estimators above.
Intuitively, this may happen because
they are local least squares estimators, and thus a few
atypical $Y_i$'s in \eqref{eq:local.const} or \eqref{eq:multiv.local.const} 
can affect the resulting $\hat{f}$ at points $\bx_0$ close to the 
$\bx_i$'s where the outliers are. 
Furthermore, the unbounded loss functions
in \eqref{eq:ise}, \eqref{eq:cv}, and \eqref{eq:mspe}, 
also imply that atypical observations 
might severly affect the selection of bandwidth parameters. 
Informally, these criteria will typically not be able to choose a 
bandwidth $h$ that predicts most of the observations very well at the 
expense of ``ignoring'' the outliers, but rather tend to favor
bandwidth values that accommodate both outliers and good data
equally mediocrely. 
In this 
section we will review 
robust local regression estimators, including  strategies to select 
the bandwidth parameter. 

Similarly to what is done in robust estimation of location or 
linear regression parameters, 
a natural approach to obtain robust estimators 
(against atypical observations in the training
set) is to replace the squared loss 
in \eqref{eq:local.const}, say, 
by one that increases more slowly, or 
even a bounded one (Maronna et al, 2018).  
Let $\rho$ be a 
loss function that is even, 
non-decreasing on $(0, \infty)$ and such that $\rho(0) = 0$. 
Typical examples
include members of the Huber or Tukey families, which are given by
\begin{equation} \label{eq:huber}
\rho^H_{c}(u) = \left\{ \begin{array}{ll} u^2  / 2 & \text{ if } |u| \le c \, , \\
c \left( |u | - c/2 \right) & \text{ if } |u| > c \, ,
\end{array} \right. \, ,
\end{equation}
and
\begin{equation} \label{eq:tukey}
\rho^T_{c}(u) = \left\{ \begin{array}{ll} 1 - \left( 1 - (u/c)^2 \right)^3 & \text{ if } |u| \le c \, , \\
1 & \text{ if } |u| > c \, ,
\end{array} \right. \, ,
\end{equation}
respectively. Note that these families are 
parametrized by a tuning constant $c > 0$. Intuitively, 
$c$ determines a threshold such that 
observations with residuals that exceed it are downweighted 
and flagged as potential outliers 
(see Maronna et al, 2018). Thus, $c$ 
should be chosen taking into account 
the ``size'' (scale) of the residuals $\varepsilon$ in \eqref{eq:model1}, 
which is typically unknown. In practice it is customary to use 
standardized residuals, and thus one can choose $c$ independently of
${\cal D}_n$ (provided a robust residual scale estimator 
is available), generally considering the asymptotic 
properties of the estimator (e.g. its asymptotic variance). 

To fix ideas, consider the 
local constant estimator with a single explanatory
variable in \eqref{eq:local.const}, with $\bX \in \calR$. 
A robust local kernel 
regression estimator of $f(x_0)$ is given by
\begin{equation} \label{eq:robloc0}
\hat{f}(x_0) = 
\argmin_{ a \in \calR} \ \sum_{i=1}^n
K_h \left( x_i - x_0  \right) \,   
\rho_c \left( \frac{ Y_i - a }{ \sigma_{x_0}} \right) \, ,
\end{equation}
where $\sigma_{x_0} > 0$ is the dispersion parameter of 
the distribution of $Y | X = x_0$ 
in \eqref{eq:cond.disp}, which in practice typically 
needs to be replaced by an estimate $\hat{\sigma}_{x_0}$. 

The estimator \eqref{eq:robloc0} for the homoscedastic 
case (where $\sigma_{x_0} = \sigma$
for all $x_0$) was studied by 
H\"ardle and Gasser (1984) and Hall and Jones (1990) 
when $\sigma$ is assumed to be known. The latter 
also proposes to select both the bandwidths and the tuning 
parameter of the loss function $\rho$ via cross validation
(see also Ma and Zhao, 2016). In addition, local linear and 
local polynomial estimators in this setting 
were studied in detail by Fan et al. (1994) and 
Jiang and Mack (2001), respectively. The latter 
consider the case of dependent 
mixing errors. 

In the case of heteroscedastic errors, H\"ardle and Tsybakov
(1988) proposed simultaneous  estimation of the regression
and scale functions using local constant kernel estimators, whereas Welsh 
(1996) studied simultaneous local polynomial estimators, 
which also yield estimators for the derivatives of the regression function. 
See also H\"ardle and Gasser (1985) and Boente 
and Rodriguez (2006) for a different approach to 
obtain robust estimators of the derivatives of the 
regression function for homoscedastic models, 
the former assuming that the common $\sigma$ 
is known. 

A natural alternative to simultaneously estimating the
regression and scale functions is to replace $\sigma_{x_0}$ in 
\eqref{eq:robloc0} by a robust scale estimator, such as a local MAD
(Boente et al., 2010). For homoscedastic models where
$\sigma_x = \sigma$, Rice (1984) proposed the estimator
$$
\hat{\sigma}_n^2 = \frac{1}{2 \, (n-1)} \sum_{i=1}^{n-1} (y_{i+1} - y_i)^2 \, .
$$
When outliers may be present in the data, 
Ghement et al. (2008) extended this approach to obtain
robust scale estimates using an  
M-scale estimator $\hat{\sigma}_n$ given by the solution to
$$
\frac{1}{n-1} \sum_{i=1}^{n-1} \rho \left( \frac{y_{i+1} - y_i }{ a \, 
\hat{\sigma}_n } \right) = b \, ,
$$
where $\rho$ is a bounded and symmetric loss function, 
and the constants $a$ and $b$ are chosen to obtain Fisher consistency. 
For example one can take 
$\rho_c$ to be a Tukey loss function as in \eqref{eq:tukey} with 
$a = \sqrt{2}$, $b = 3/4$ and $c = 0.70417$.

%
%



Note that robust estimators of the form \eqref{eq:robloc0} may be
thought of as ``plug in'' approaches, 
in the sense that they are often proposed as a ``natural'' modification 
of the ``classical estimator''. Specifically, the approach in
\eqref{eq:robloc0} replaces the square loss function in 
\eqref{eq:local.const} with the function $\rho_c$, but
the reasoning leading to \eqref{eq:local.const} relies on 
the assumption that 
the errors 
satisfy $E[ \varepsilon | X = x] = 0$ 
for (almost) all 
$x$, and consequently  that
$f(x) = E[ Y | X = x]$, which motivates \eqref{eq:nadaraya}. 
However, to allow for heavy tailed error distributions 
one would like to avoid assumptions (explicit or implicit) 
on the moments of the errors in \eqref{eq:model1}. 
In other words, it may not be immediately clear what function 
the estimator in 
\eqref{eq:robloc0} is estimating 
if the errors $\varepsilon$ do not have finite moments. 
Boente and Fraiman (1989a) defined a type of ``robust conditional 
expectation'' (called ``robust conditional location functional'') 
and showed that, if $\psi$ is the derivative of a convex 
loss function $\rho$
then the  robust conditional location functional 
is characterized by being the (unique) measurable function $g^\psi$ such that 
$$
E \left[ h(X) \psi \left( \frac{Y - g^\psi(X)}{\sigma} \right) \right]  = 0 \, ,
$$
for any integrable function $h$. 
An estimator $\hat{g}^\psi(x)$ of $g^\psi(x)$ is given by the solution of 
\begin{equation} \label{eq:robust.cond}
E_{\hat{F}_{Y|X=x}}  \left[ \psi \left( \frac{Y - \hat{g}^\psi(x)}{\sigma} \right) \right]  = 0 
\end{equation}
where $\hat{F}_{Y|X=x}$ is an estimator of the conditional distribution of $Y | X = x$. 
Boente and Fraiman (1989b) and (1990) extended this approach to cases with 
dependent (mixing) errors.  We can use the results in 
Boente and Fraiman (1989a) to show that 
\eqref{eq:robloc0} is the equivalent of 
solving \eqref{eq:robust.cond}
with a kernel density estimator
$\hat{F}_{Y|X=x}$.

\subsubsection{Robust bandwidth selection} \label{sec:robust.bandwidth}

A critical step in computing a robust nonparametric regression estimator
is the selection of the bandwidth $h$ in \eqref{eq:robloc0}. 
There are (just like for the non-robust methods) two approaches to 
choose a value for $h$: ``plug in'' estimators, or some variant of cross-validation.  
For local constant M-type kernel estimators for a homoscedastic model
with known error scale, 
optimal plug-in methods to select $h$ 
were studied by H\"ardle and Gasser (1984) and
Hall and Jones (1990). For local linear and local polynomial estimators 
plug-in bandwidth selection was studied 
by Fan et al (1994) and Jiang and Mack (2001) respectively. 
When the error scale needs to be estimated (as is usually the case in practice)
plug-in optimal bandwidths for local constant estimators 
were proposed by Boente et al. (1997). 
These plug-in estimates for the optimal bandwidth are of the form:
$$
\hat{h}_n \, = \, n^{-1/5} \, \left( \frac{ \hat{\sigma}^2 \, C_K }{ \int (\hat{f}''(u))^2 du } \right)^{1/5} \, ,
$$
where $C_K = \int K^2(u) du / \left( \int u^2 K(u) du \right)^2$. 
Hence note that in order to use these bandwidth estimates in practice 
one needs to have reliable estimators $\hat{\sigma}$
and $\hat{f}''$ for the error scale $\sigma$  and
the second derivative of the regression function, respectively. The latter 
typically has to rely on a preliminary or pilot bandwidth. 

Although cross-validation is an attractive practical alternative for 
bandwidth selection, 
as it was noted by several 
authors, when the data may contain outliers one needs to 
make sure the CV criterion 
does not ``penalize'' candidate bandwidths that result in fits that
predict well most of the observations while not adjusting a small proportion of
them. Thus, the CV objective function 
needs to be robust itself. 
Wang and Scott (1994) working with an $L_1$ 
local regression estimator proposed using a corresponding $L_1$ cross
validation criterion: 
\begin{equation} \notag
	\text{RCV}_{L_1}(h) \, = \, \frac{1}{n} \sum_{i=1}^n \bigl| 
y_i - \hat{f}_h^{(-i)}(x_i) \bigr| \, .
\end{equation}
More generally, it is natural to consider replacing 
the squared prediction errors in the usual cross-validation criterion by a
function $\rho$, as it is done with the estimating equations:
\begin{equation} \label{eq:rcv}
\text{RCV}_\rho(h) \, = \, \frac{1}{n} \sum_{i=1}^n \rho \left( 
\frac{y_i - \hat{f}_h^{(-i)}(x_i)}{\hat{\sigma}_n} \right) \, ,
\end{equation}
where $\rho$ need not be the same one used to compute the robust
regression estimate $\hat{f}$. 
For homoscedastic models with known error scale, Leung (2005) 
(see also Leung et al., 1993) showed that such a robust 
cross validation method is asymptotically correct.
To select the 
tuning parameter of the $\rho$ function in \eqref{eq:rcv}
one needs to compute $\hat{\sigma}_n$, an estimate of the error scale $\sigma$.
For homoscedastic models, one can use a robust scale estimator of
the residuals obtained with a preliminary regression estimator 
(e.g. one obtained by selecting the bandwidth via an $L_1$  CV
criterion) (Hogg, 1979), or use a direct preliminary robust error scale estimate
as in Ghement et al. (2008). 
Lee and Cox (2010) reported that the performance of these
alternatives are very similar to each other. 


Note that the classical cross validation criterion \eqref{eq:cv} can be
written as the sum of the sample variance of the prediction residuals
plus the square of their sample mean
$$
\text{CV}(h) \, = \, \bar{e}_n^2 + \frac{1}{n} \sum_{i=1}^n 
(e_i - \bar{e}_n )^2 \, ,
$$
where
$e_j = y_j - \hat{f}_h^{(-j)}(x_j)$, $1 \le j \le n$, and 
$\bar{e}_n = \sum_{i=1}^n e_i / n$ (note that the $e_j$'s depend 
on $h$, but we ignore this to keep our notation slightly simpler). 
Thus, a natural ``plug in'' robust alternative to $\mbox{CV}(h)$ is 
given by 
\begin{equation} \label{eq:rcv2}
\text{RCV}_2(h) \, = \, \hat{\mu}_n^2(e_1, \ldots, e_n) + \hat{\sigma}_n^2(e_1, \ldots, e_n) \, .
\end{equation}
where $\hat{\mu}_n$ is a robust location estimator of the prediction residuals
$e_1, \ldots, e_n$, and $\hat{\sigma}_n$ is a robust scale estimator of the same
residuals (see e.g. Bianco
and Boente, 2007; Boente and Rodriguez, 2008; and Boente et al. 2012).
Although no theoretical results exist to justify selecting $h$ as the minimum of 
$\mbox{RCV}_2$ above, such a bandwidth choice strategy 
generally works well in numerical experiments.

\subsection{Splines and basis expansions} \label{sec:splines}

A different approach for constructing nonparametric regression estimators 
is to represent $f$ on a basis $\phi_1$, $\ldots$, $\phi_{k_n}$,
with $k_n \to \infty$ as $n \to \infty$. 
Although several different bases can be used
(e.g. Fourier, polynomial, etc.), spline bases are 
probably the most popular
in the Statistics literature. 
Splines are smooth functions of the form 
$$
\phi_j(x) = \left[ \left(x - \kappa_j \right)_+ \right]^p \, , \quad
x \in \calR \, ,
$$
where $a_+ = a I(a \ge 0)$, $p \in \mathbb{N}$, and 
$\kappa_j \in \calR$ are known as the knots. Thus, 
to use a splines basis we need to specify its degree $p$ 
and the number and value of the knots $\kappa_1$, $\ldots$, 
$\kappa_{k_n}$. In addition to the $\phi_j$'s above, the 
spline basis
also includes the functions $1$, $x$, $x^2$, $\ldots$, $x^p$. 
Once the basis is chosen, 
to find the best function in the 
corresponding linear subspace 
we need to solve
\begin{equation} \label{eq:reg.splines}
\min_{\bbeta \in \calR^K} \sum_{i=1}^n \left( y_i - 
\bbeta^\top \Gamma_i \right)^2
\end{equation}
where $\Gamma_i = (\gamma_{1}(x_i), \ldots, \gamma_{K_n}(x_i))^\top \in 
\calR^{K_n}$ 
and $\gamma_\ell(x)$ are the functions in the spline basis. 
Hence, in this case
the non-parametric
regression estimation problem reduces to a standard linear regression
one. 
Note that although the basis as discussed here is easy to describe and 
understand, it may be numerically rather unstable, and thus one
might prefer to use an alternative one that spans the same 
subspace but has better computational properties. 
B-spline bases are such an option, and thus typically preferred in 
practice to the simpler truncated polynomial version above. 



Naming conventions vary, but we can nevertheless
identify three conceptually different approaches using 
bases to represent the regression function $f$:
regression splines, penalized regression splines and smoothing splines. 
Regression splines estimators are constructed by judiciously choosing 
the number and
position of the knots, and computing the corresponding approximation on 
the subspace generated by the resulting basis as in \eqref{eq:reg.splines}.
Having chosen the basis,
this reduces to estimating the coefficients of a linear regression model. 

Penalized regression splines avoid the computationally costly 
problem of objectively selecting the number and location of the knots by 
using a relatively large number of them (but less than 
the size of the training set), often equispaced, 
plus a penalty directly expressed in terms of the spline coefficients 
(Eilers and Marx, 1996; Ruppert et al., 2003; 
Xiao, 2019). They solve 
\begin{equation} \label{eq:pen.splines}
\min_{\bbeta \in \calR^K} \sum_{i=1}^n \left( y_i - 
\bbeta^\top \Gamma_i \right)^2 \, + \, P_\lambda ( \bbeta ) \, ,
\end{equation}
where $P_\lambda ( \bbeta )$ is a penalty term, and $\lambda > 0$ is
a smoothing coefficient. For example we may have 
$P_\lambda(\bbeta) = \lambda \, \| \bbeta \|_2^2$, or 
$P_\lambda(\bbeta) = \lambda \, \sum_{j=2}^{K_n} (\beta_j - \beta_{j-1})^2$.  

Finally, smoothing splines are defined directly as the minimizer of 
\begin{equation} \label{eq:smoothing.spline}
\sum_{i=1}^n (y_i - h(x_i))^2 + \lambda \, \int (h''(t))^2 dt
\end{equation}
over all functions $h$ with absolutely continuous first derivative. 
It can be shown that the minimizer $\hat{h}$ 
is a \textit{natural} cubic spline, that is: a spline with $p=3$, 
knots at each of the observed $x_i$'s, and
with the added restriction that it will be 
a linear function before the smallest knot and after the largest one.
Such splines are called natural cubic splines. 
We refer the reader to Green and Silverman (1994) 
and Ruppert et al. (2003)
for more details. 

Most of the literature on splines deals with models with single
explanatory variables, although splines bases for multiple
predictors have been proposed (see, e.g., Cox, 1984; and 
Gy\"orfi et al., 2010).  In particular, thin-plate 
and tensor product bases can be constructed for any number
of explanatory variables. The main challenge when using 
the former is their
computational cost, and for the latter the difficulty lies in 
quantifying (and thus, efficiently penalizing) smoothness
(see Wood, 2006, Section 4.1). 


\subsubsection{Robust splines estimators}

The first robust proposals for 
this type of approach considered smoothing splines
(Huber, 1979). 
The idea is to minimize 
\eqref{eq:smoothing.spline} but using a robust loss 
function $\rho_c$ instead: 
\begin{equation} \label{eq:robust.smoothing.spline}
\sum_{i=1}^n \rho_c \left( \frac{y_i - h(x_i)}{\hat{\sigma}}
\right) + \lambda \, \int (h''(t))^2 dt \, ,
\end{equation}
where $\hat{\sigma}$ is a robust residual scale estimator. 
Note that, as discussed before, 
we need an estimate of the residuals scale $\hat{\sigma}$, which should be
robust, and also computed separately from $\hat{f}$. Cantoni and Ronchetti (2001)
suggest using robust variants of the proposals in 
Gasser et al. (1986) and Cunningham et al., (1991), but one could also
use the robust scale estimator in Ghement et al. (2008), for example. 
When $\rho_c$ is convex (as in Huber's family \eqref{eq:huber}),
the minimizer of \eqref{eq:robust.smoothing.spline} is a natural cubic spline, and
the solution $\hat{\mathbf{f}} = (\hat{f}(x_1), \hat{f}(x_2), \ldots, 
\hat{f}(x_n))^\top$ satisfies the following system of equations:
$$
-\psi( \mathbf{r} ) + 2 \, \lambda \, \mathbf{B} \, \hat{\mathbf{f}} = \mathbf{0} \, ,
$$
where $\psi( \mathbf{r} ) = (\psi(r_1), \psi(r_2), \ldots, \psi(r_n))^\top$, 
$r_i = (y_i - \hat{f}(x_i))/\hat{\sigma}$, 
$\mathbf{B} = \Gamma \, \mathbf{D} \, \Gamma$, $\Gamma$ is the 
natural cubic spline matrix, and $\mathbf{D}$ is the matrix of integrated second 
derivatives of the basis. 
Cox (1983) provided a thorough theoretical treatment
of M-type smoothing splines for homoscedastic models with known 
error scale. Cunningham et al. (1991) studied the case of unknown 
error scale. 
See also Oh et al. (2007) for further discussions on the role
of pseudo observations in robust smoothing. 
More recently, Kalogridis (2022a) provided a unified treatment of 
M-type smoothing splines, 
and Kalogridis (2022b) studied robust M-estimators based on 
multivariate thin-plate splines for 
the case where the error scale $\sigma$ is known.

Robust M-type regression splines estimators were
studied by Shi and Li (1995) and He and Shi (1995) for models
with known and unknown error scales, respectively. 
The idea is to replace the squared loss in \eqref{eq:reg.splines} with a
robust one, like Huber's in \eqref{eq:huber} and solve
\begin{equation} \label{eq:robust.reg.splines}
\min_{\bbeta \in \calR^K} \sum_{i=1}^n \rho_c \left( \frac{y_i - 
\bbeta^\top \Gamma_i}{\hat{\sigma}_n} \right) \, , 
\end{equation}
where the residual scale estimate $\hat{\sigma}_n$ needs to be computed
separately (based on convergence considerations, the authors suggest following
Gasser et al., 1986; and Cunningham et al., 1991; other approaches
may work as well).  Shi (1997)
studied regression splines for dependent data when the error
scale is known. The number and location of the knots 
should be chosen carefully, taking into account the potential
issues caused by outliers in the training set, which results in a 
complex optimization problem that is generally computationally 
very costly. 


The literature on robust penalized regression splines is less abundant. 
Lee and Oh (2007) studied 
M-type penalized regression spline estimators for homoscedastic models with 
unknown error scale. 
They
proposed an iterative algorithm based on so called ``pseudo data'',
which updates the error scale estimator in each iteration. 
A family of robust penalized regression spline estimators 
for the same models with homoscedastic errors and unknown scale
that do not
require the preliminary estimation of the error scale is given by the S-estimators
of Tharmaratnam et al. (2010). 
Specifically, they proposed to solve
\begin{equation} \label{eq:s.pen.splines}
\min_{\bbeta \in \calR^K} \, \left[ n \, \hat{\sigma}^2_n(\bbeta) + 
P_\lambda( \bbeta) \right] \, , 
\end{equation}
where $P_\lambda( \bbeta)$ is the penalty term, and $\hat{\sigma}_n(\bbeta)$ is a 
robust scale estimator of the residuals, for example, an M-scale solving
$$
\frac{1}{n} \sum_{i=1}^n \rho \left( \frac{y_i - 
\bbeta^\top \Gamma_i}{\hat{\sigma}_n(\bbeta)} \right) = b \, ,
$$
where the constant $b \in \calR$ is chosen such that $E_{F_0} [ \rho(U) ] = b$ (which results
in Fisher consistency) and $\rho$ is bounded and symmetric around zero (e.g. one of Tukey's 
loss functions in \eqref{eq:tukey}). 
When the penalty $P_\lambda(\bbeta) = \bbeta^\top \, \mathbf{D} 
\, \bbeta$, for a fixed matrix $\mathbf{D}$, 
Tharmaratnam et al. (2010) showed that the estimator can be computed using 
a simple iterative algorithm. Similarly to what is done for linear
regression models, one can use these S-penalized regression spline estimators,
and their associated error scale estimator, as a starting 
point to compute M-type penalized regression splines. 
The numerical experiments reported by Wang et al (2014)
indicate that the S-estimators perform very well in a wide range of
settings, at the expense of an increased computational budget.  
A thorough study of M-type penalized 
regression spline 
estimators 
(and algorithms to compute them) for homoscedastic nonparametric regression
models with unknown error scale can be found in 
Kalogridis and van Aelst (2021 and 2022). The estimators solve the natural
robust extension of \eqref{eq:pen.splines}:
\begin{equation} \label{eq:robust.pen.splines}
\min_{\bbeta \in \calR^K} \sum_{i=1}^n \rho_c \left( \frac{ y_i - 
\bbeta^\top \Gamma_i}{\hat{\sigma}_n} \right) \, + \, P_\lambda ( \bbeta ) \, ,
\end{equation}
where $\hat{\sigma}_n$ is a robust residual scale estimator, computed independently,
as in Ghement et al. (2008). 

%
%

%

\subsubsection{Robust selection of the penalty parameter}

The selection of the regularization parameter 
for penalized regression splines estimators is typically done using
some variant of cross validation, and it presents the same challenges found when 
choosing bandwidths for local kernel estimators (Section \ref{sec:robust.bandwidth}),  
namely: that the cross validation 
criteria need to be robust. Lee and Oh (2007) proposed to update the 
penalty parameter at each iteration of their algorithm, 
using a standard generalized cross-validation (GCV) criterion. 
The main idea behind GCV is to note that, for linear estimators (as most of these 
penalized regression splines ones) 
there is a closed form expression for the leave-one-out CV criterion
\eqref{eq:cv} that avoids having to re-fit the model $n$ times. Specifically,
$$
\text{CV}(\lambda) = \frac{1}{n} \sum_{i=1}^n \left( \frac{y_i - 
\hat{\bbeta}_n^\top \Gamma_i}{ 1 - \mathbf{H}_{ii}(\lambda) } \right)^2 \, ,
$$
where $\mathbf{H}_{ii}(\lambda)$ is the $i$-th element on the diagonal of the
``hat'' matrix
corresponding to the solution of \eqref{eq:pen.splines}. 
However, sometimes it is computationally simpler to compute the
trace of $\mathbf{H}(\lambda)$ than its individual terms, and thus the 
Generalized Cross-Validation criterion replaces each
$1 - \mathbf{H}_{ii}(\lambda)$ 
above by their average $1 - \text{trace}(\mathbf{H}(\lambda)) / n$:
\begin{equation} \label{eq:gcv}
\text{GCV}(\lambda) = n \, \sum_{i=1}^n \left( \frac{y_i - 
\hat{\bbeta}_n^\top \Gamma_i}{ n - \text{trace}(\mathbf{H}(\lambda)) } \right)^2 \, .
\end{equation}
See, e.g. Ruppert et al. (2003) and references therein for more details. 

As mentioned above, using a criterion as \eqref{eq:gcv} when outliers
may be present in the data can result in a non-robust fit, even if one uses
a robust estimator to compute $\hat{\bbeta}_n$. 
The estimators in \eqref{eq:s.pen.splines} and 
\eqref{eq:robust.pen.splines} can be calculated using iterative re-weighted 
least squares, and thus, at convergence the robust estimate
$\hat{\bbeta}_n$ is the solution of a weighted least squares
penalized spline problem (but note that the weights depend on $\hat{\bbeta}_n$). 
This suggests using a weighted version of 
\eqref{eq:gcv}, where the weights $w_i(\hat{\bbeta}_n) = 
\rho'_c(r_i) / r_i$, $1 \le i \le n$, $r_i = (y_i - 
\hat{\bbeta}_n^\top \Gamma_i) / \hat{\sigma}_n$, and $\rho'_c(u)$
denotes the derivative of the loss function.
Tharmaratnam et al. (2010)
and Kalogridis and van Aelst (2021, 2022) recommended using the
corresponding weighted version of \eqref{eq:gcv}:
\begin{equation} \label{eq:wgcv}
\text{GCV}_W(\lambda) = \frac{ n \, \sum_{i=1}^n w_i(\hat{\bbeta}_n) ( y_i - 
\hat{\bbeta}_n^\top \Gamma_i )^2 }{ ( n - \text{trace}(\mathbf{H}_W(\lambda)) )^2 } \, ,
\end{equation}
where $\mathbf{H}_W$ is the ``hat'' matrix that corresponds to the weighted
least squares problem: $\mathbf{W}^{1/2} \mathbf{H} \mathbf{W}$, with 
$\mathbf{W} = \text{diag}(w_1, 
\ldots, w_n)$. 
Since several of these weights may be zero, 
Tharmaratnam et al. (2010)  replaced
$n$ in \eqref{eq:wgcv} with the number of 
non-zero weights, $n_w$, say. 
Minimizing $\text{GCV}_W$ over $\lambda$ typically requires 
evaluating it over a relatively fine grid of possible values of the regularization parameter. 

Robust selection of the penalization parameter for M-type smoothing 
splines was discussed by Cantoni and Ronchetti (2001). They proposed
both a robust version of Mallows's $C_p$, and a closed-form expression
for a robust leave-one-out cross-validation criterion. The robust $C_p$
criterion is given by 
$$
\text{RC}_p(\lambda) = \sum_{i=1}^n \psi_c^2 \left( \frac{y_i - \hat{f}_\lambda(x_i) }{ \hat{\sigma} }
\right) + A_n \, ,
$$
where $\psi_c = \rho_c'$, 
and $A_n$ is a correction term to make $\text{RC}_p$ an unbiased 
estimator of the (weighted) prediction error (Ronchetti and Staudte, 1994). 
This correction depends on $\rho_c'$ and $\rho_c''$. 
Cantoni and Ronchetti (2001) also discussed a simpler criterion 
to select the penalization parameter for robust smoothing
splines, which is a closed-form expression for a robust version of 
leave-one-out cross validation. Namely:
$$
\text{RCV}(\lambda) =  \frac{1}{n} \frac{\hat{\sigma}^2}{\overline{\Psi'}_{cn}^2} 
\sum_{i=1}^n \left( \frac{ \psi_c( r_i) }{ 1 - S_{ii} }
\right)^2 \, ,
$$
where $\overline{\Psi'}_{cn} = (1/n) \sum_{i=1}^n \psi_c'(r_i)$ and 
$r_i = (y_i - \hat{f}_\lambda(x_i)/\hat{\sigma}$, and $S_{ii}$ are the diagonal
elements of the ``hat'' matrix associated with the ``pseudo observations''
representation of the robust smoother (Cox, 1983). 


%
%
%
%

\section{Semi-structured methods}

Applying the methods discussed in Section \ref{sec:unstr}
to models with a relatively large number of 
explanatory variables presents several conceptual and practical 
difficulties.  For example, local kernel
estimators can be seriously affected by the well-known  
``curse of dimensionality'' (Stone, 1980): 
as the number of variables increases,
local estimation methods 
require exponentially larger training sets to maintain their 
accuracy. Spline-based estimators, for their part, become
challenging because constructing high-dimensional 
multivariate spline bases can be computationally very costly
(Wood, 2006, Section 4.1). 

In this section we discuss three approaches that can be used
to estimate the regression function 
when $\bX$ contains many explanatory variables. 
Additive models, although still nonparametric, incorporate a relatively rigid
structure in the regression function, and manage to avoid the 
``curse of
dimensionality'' 
and the need to construct multivariate spline bases. 
Regression trees approximate $f$ in \eqref{eq:model1}
with a function that is constant over disjoint regions ${\cal R}_j$. 
These constants and regions are approximated using a (greedy) 
recursive partitioning algorithm. Finally, 
boosting methods combine a large number of 
simple functions (e.g. regression trees with one or two splits)
to estimate $f$. These simple functions can be computed 
very efficiently even with a large number of available explanatory
variables.  

\subsection{Additive models}

Imposing an additive structure to the regression
function $f$ allows us to obtain
estimators that can avoid being affected by the 
``curse of dimensionality''. 
Notably, Stone (1986) showed that using splines to estimate 
these additive components 
results in estimators with optimal one-dimensional rates of convergence.
Additive models 
postulate that the regression function $f$ in \eqref{eq:model1}
satisfies
\begin{equation} \label{eq:am1}
f( \bX ) = \alpha + \sum_{j=1}^d f_j(X_j) \, ,
\end{equation}
where $\alpha \in \calR$, and the $f_j$'s are unspecified smooth functions of 
each explanatory variable. In order for the components of the above model
to be identifiable one usually assumes further that $E[ f_j(X_j) ] = 0$ for 
$j = 1, \ldots, d$.  
The idea of imposing an additive structure to the regression function
is present, for example, behind the projection pursuit model of Friedman and Stueztle
(1981). Hastie and Tibshirani (1990) provide a comprehensive
discussion of additive models. 

One relatively straightforward method to estimate the additive 
components of the model above is to use splines. Specifically,
one can represent each $f_j$ in \eqref{eq:am1} 
using a spline basis (possibly different bases for different
$f_j$'s), and reformulate the problem into
the framework of Section \ref{sec:splines}. 
Robust estimators can be obtained following the same ideas
discussed above for spline methods
(see, e.g., Boente et al., 2020; Boente and Mart\'\i nez, 2023, 
where the non-parametric components are treated in this way). 
Note that with this approach there are $d$ penalty parameters to be selected
(one for each $f_j$), which can be a very 
challenging problem for moderate or large values of $d$. 
For classical penalized spline estimators, 
Wood (2000) and Wood et al (2016) developed a computationally
feasible way to numerically minimize the generalized 
cross validation criterion without having to explore
a $d$-dimensional grid. As far as the author knows, no such general and 
feasible strategy exists for robust estimators. 

Another approach to estimate the $f_j$'s in \eqref{eq:am1} is the
backfitting algorithm (Friedman and Stueztle, 1981; Buja et al. 1989). 
This method is based on the observation that \eqref{eq:am1} 
and \eqref{eq:model1}, together
with $E(\varepsilon | \bX = \bx) = 0$ for (almost) all $\bx$, imply
that the additive components satisfy
\begin{equation} \label{eq:backfit}
	f_\ell(X_\ell) = E \Bigl[ \Bigl. Y - \alpha - 
	\sum_{j \ne \ell}^d f_j(X_j) \Bigr| X_\ell \Bigr]
	\, , \quad \ell = 1, \ldots, d \, ,
\end{equation}
which suggests an iterative algorithm where each $f_\ell$ 
is estimated with a univariate smoother of the partial 
residuals $Y - \alpha - \sum_{j \ne \ell} f_j(X_j)$ as a function of
$X_\ell$. Starting with initial estimates 
$f_j^{(0)}$, $j = 1, \ldots, d$, we compute the
$f_\ell^{(1)}$'s by using a non-parametric regression
estimator for the partial residuals 
as a function of the $\ell$-th explanatory variable. 
After we obtain all the updated
$f_\ell^{(1)}$'s  the process is iterated until convergence. 
Intuitively, one can obtain a robust version of the
backfitting algorithm by using any robust univariate
non-parametric regression estimator in the algorithm
above. 
Boente et al. (2017) showed that such 
a robust backfitting algorithm based on 
M-type robust kernel local regression estimators
actually finds the functions $f_j$ and 
parameter $\alpha$ that minimize
$$
E \left[ \rho \left( \frac{ Y - \alpha - \sum_{j=1}^d f_j(X_j) }{
	\sigma} \right) \right] \, , 
$$
where $\rho$ is a robust loss as before,
and $\sigma$ is the dispersion parameter of the errors. This allowed
Boente et al. (2017) to show that this approach provides 
consistent and robust estimators of the components of \eqref{eq:am1}. 
Bianco and Boente (1998) studied a related approach that
applies to models satisfying a relatively 
restrictive assumption.
Note that the backfitting algorithm requires selecting 
$d$ bandwidth parameters, which can be computationally
costly. 
To the best of our knowledge, the only option for finding
the bandwidths is 
brute-force minimization of a robust 
measure of prediction error, such as \eqref{eq:rcv2}. This 
optimization is carried out by evaluating the objective
function pointwise over a relatively fine
$d$-dimensional grid of possible bandwidths. 
Finally, Boente and Martinez (2017) proposed a robust
version of 
the marginal integration approach to estimate the components
of additive models
(Tj\o stheim and Auestad, 1994; and Linton and Nielsen, 1995). 


\subsection{Regression trees and Random Forests}

Regression trees were introduced by Breiman et al. (1984). 
They estimate the regression function $f(\bX)$ with  
one that is constant over a set of disjoint regions of the domain: 
$$
g( \bX ) = \sum_{j=1}^K a_j \, I( \bX \in {\cal R}_j ) \, ,
$$
where $a_j \in \calR$ and ${\cal R}_j \subset \calR^d$. 
Finding optimal $a_j$'s and ${\cal R}_j$'s is an extremely 
challenging computational problem. A greedy algorithm
approximates the function $g$ with 
sets ${\cal R}_j$ that are found by iteratively splitting the
space into half-planes  of the form
$\left\{ \bX \in \calR^d : X_\ell \le b \right\}$  and
$\left\{ \bX \in \calR^d : X_\ell > b \right\}$, 
for some 
$1 \le \ell \le d$ and $b \in \calR$. These cuts, and
the constant $b \in \calR$ are chosen to
reduce the squared residual error the most at each split. 
The resulting $g$ can also be efficiently described using
a binary tree. 

Random Forests 
(Breiman, 2001) are ``bagging''-type ensembles (Hastie et al., 2009) 
of regression trees, 
but constructed in a specific way designed 
to reduce the ``correlation'' between the different trees in the
ensemble (the ``forest''). This is done by only considering a
randomly selected 
subset of variables as possible splits for each leaf.  
This correlation reduction generally results in predictions with 
notably less variability and thus improved squared prediction errors. 

A relatively small proportion of observations
with atypical values in the response variable may 
affect noticeably both regression trees and random forests. A natural
approach to construct robust alternatives to the algorithm 
above is to replace the squared residual error that is minized at 
each split by 
a different loss measure (e.g. absolute values, or Huber's $\rho$).
Numerical experiments typically show that 
$L_1$-trees (Breiman et al, 1984)
can perform better than the standard 
$L_2$ ones when a small proportion of moderate outliers 
is present in the training set.  
Galimberti et al (2007) studied using the Huber's loss function to
construct regression trees (see also Chambers et al., 2004). 
However, this strategy 
will require that the error scale be re-estimated 
for each leaf to compute the reduction in residual ``loss'', 
which necessarily changes the loss function 
that is being optimized at 
each split. As a consequence, 
if we use a different scale in each ``child'' leaf,
we may not be able to find a split that 
reduces the overall loss. Alternatively, if 
we use the scale estimator of the ``parent'' leaf when we
compute the ``loss'' of each child leaf, the loss function
changes at each iteration. The fact that it is not clear whether
a global loss criterion exists that is being minimized by the algorithm
makes it difficult to study the properties of this type of methods. 
A different approach to 
dealing with potential outliers when building trees was given by 
John (1995), where the concept of ``leverage points'' (those observations
exerting an undue influence on the resulting fit) is extended to 
binary trees. Unfortunately, as the number of explanatory variables 
increases this approach quickly becomes computationally 
too costly to be feasible. 

\subsection{Boosting}

Gradient boosting (Friedman, 2001) proposed a different type of
ensemble estimators for regression that is constructed
sequentially. The idea is borrowed from the standard gradient search
algorithm for optimization. Given a loss function $L : \calR^2
\to \calR_+$, define the objective regression function $f$
as 
$$
f = \arg \min_{g \in {\cal G}} E \left[ L \left( Y, g(\bX)
\right) \right] \, ,
$$
where ${\cal G}$ denotes a suitably defined class of functions.
The empirical version of the above population problem is 
$$
\hat{f}_n = \arg \min_{g \in {\cal G}} 
\frac{1}{n} \sum_{i=1}^n L \left( Y_i, g(\bX_i)
\right) \, .
$$
Since in general one cannot calculate $\hat{f}_n$ above directly, we 
approximate it using
the following iterative algorithm. Let $\hat{f}^{(k)}$ denote the 
approximation at step $k$, then
$$
\hat{f}^{(k+1)} = \hat{f}^{(k)} + \gamma \, \hat{\alpha}^{(k+1)} \, 
\hat{h}_{k+1} \, , 
$$
where $\hat{h}_{k+1}$ is a ``weak'' approximation to the
negative gradient of the empirical risk at step $k$. For example,
$\hat{h}_{k+1}$ can be a regression tree with only one or two
splits fitting the $n$ negative gradient values
$$
u_i^{(k)} = \left. \frac{\partial }{\partial b}L(Y_i, b) \right|_{b = 
\hat{f}^{(k)}(\bX_i)} \, , \qquad i=1, \ldots, n, \, ,
$$
where $\hat{\alpha}^{(k+1)}$ is the step size of a line
minimization along the ``direction'' given by 
$\hat{h}_{k+1}$:
$$
\hat{\alpha}^{(k+1)} = \arg \min_{a \in \calR} 
\frac{1}{n} \sum_{i=1}^n L \left( Y_i, \hat{f}^{(k)}(\bX_i)
+ a \, \hat{h}_{k+1}(\bX_i) 
\right) \, .
$$
This general formulation of the method suggests that in order
to obtain a robust version of this approach one can take 
$L$ to be a ``robust loss'', for example Huber's function 
(Friedman, 2001). However, as mentioned before, in order to 
select the tuning constant of the Huber loss function, one needs
to estimate (robustly) the error scale, which typically requires
residuals from a robust fit. Friedman (2001) suggests using 
a different residual scale estimator at each iteration, but this
approach may not work well in practice. A possible explanation is
that, as with the approach of Galimberti et al. (2007), 
the objective function changes in each iteration. 
In other words, by changing the residual scale estimator, the gradients
computed at each iteration correspond to different loss functions.
Moreover, unduly large residual scale estimators 
(which are not uncommon in early iterations) may mask atypical
observations and result in a non-robust fit. 

An alternative is to use the $L_1$ loss, which does 
not require standardized residuals. Experiments
reported in Ju and Salibian-Barrera (2021) showed that this approach
may not work well with large outliers. Furthermore,
as we observe in other settings, a properly defined M-type
fit typically results in predictions with markedly 
better properties. 
Lutz et al (2008) proposed a robust variant
of their $L_2$ boosting approach, which uses simple 
linear regression (using a single explanatory 
variable) as weak learners. 

In order to have a robust version of gradient boosting that 
is defined from first principles (in the sense of its iterations corresponding to
a gradient descent-like algorithm in functional space, as 
originally proposed by Friedman, 2001), Ju and Salibian Barrera (2021) 
proposed to 
use the following objective function
\begin{equation} \label{eq:s-boost}
\hat{f}_n = \arg \min_{g \in {\cal G}} \hat{\sigma}_n( g ) \, ,
\end{equation}
where $\sigma(g)$ is a robust scale estimator of the residuals $Y - g(\bX)$, 
for example, an M-scale estimator:
$$
\frac{1}{n} \sum_{i=1}^n  \rho_0 \left( \frac{Y_i - g(\bX_i)}{\hat{\sigma}_n(g)} \right)  = 1/2 \, .
$$
We can compute the necessary gradients of the implicitly defined 
function $\hat{\sigma}_n( \cdot )$ and use the general formulation of
gradient boosting as proposed by Friedman (2001). Furthermore, the resulting
estimator $\hat{f}_n$ 
can be used as an initial fit to solve 
\begin{equation} \label{eq:m-boost}
\tilde{f}_n = \arg \min_{g \in {\cal G}} \sum_{i=1}^n  \rho \left( \frac{Y_i - g(\bX_i)}{\hat{\sigma}_n(\hat{f}_n)} \right) \, ,
\end{equation}
where the error scale estimator $\hat{\sigma}_n( \hat{f}_n)$ is 
the one obtained from \eqref{eq:s-boost} and kept fixed over the 
iterations to approximate $\tilde{f}_n$, the solution to \eqref{eq:m-boost}. 
Unlike the methods discussed in Section 2, this algorithm scales very well
with the number of explanatory variables and can be computed efficiently 
even with a large number of them. Extensive simulation studies 
reported in Ju and Salibian-Barrera (2021) 
show that this
robust boosting regression estimator is robust against the presence of different
types of outliers, and compares very well with existing robust proposals, 
and also with non-robust proposals when no
outliers are present in the training set. 


\section{Example}

We will illustrate the use of some of the different approaches discussed above 
by applying them to the ``Babinet'' data
(Cleveland, 1993). The data are available in the object
\texttt{polarization} of the \texttt{ggcleveland} in \texttt{R} 
(R Core Team, 2022). The data were collected 
from an experiment on the scattering of sunlight in the atmosphere. The 
response variable is the scattering angle where polarization dissappears 
(the Babinet point), and the explanatory variable is the cubic root of the
particulate concentration of a gas in the atmosphere. To avoid 
numerical issues with repeated values, this variable was perturbed 
by adding a small amount of random uniformly distributed noise. The variable
takes values between 2 and 5, and the mean absolute difference between the original 
and the perturbed variable is less than 0.008. There are $n = 355$ observations. 

We compare 5 non-parametric regression estimators: 
\begin{itemize}
\item \texttt{mgcv}: the usual smoothing spline estimator, with penalization selected 
via GCV (Wood, 2006); 
\item \texttt{M-sm1}: M-smoothing splines (Kalogridis and Van Aelst, 2021); 
\item \texttt{M-sm2}: M-smoothing splines, computed as in Oh et al. (2004); 
\item \texttt{S-pen}: S-penalized splines (Tharmaratnam et al., 2010); and 
\item \texttt{M-loc}: M-local linear regression (Boente et al. 2017). 
\end{itemize}
The author would like to thank these authors for making their code publicly available. 
\texttt{mgcv} is available in the \texttt{mgcv} package on CRAN; 
\texttt{M-sm1} is available at \newline
\texttt{https://github.com/ioanniskalogridis/Smoothing-splines}; 
\texttt{M-sm2} is implemented in the \texttt{fields} package on CRAN; 
\texttt{S-pen} is available at \newline
\texttt{https://github.com/msalibian/PenalizedS}; and 
\texttt{M-loc} is available in the \texttt{RBF} package on CRAN. 

Although both \texttt{M-sm1} and \texttt{M-sm2} are M-type smoothing splines,
they are computed differently (the latter using so-called ``pseudo-data''), and
they employ different variants of GCV to select the penalization parameter. 
\texttt{M-sm1} (like \texttt{S-pen}) uses the weighted version of GCV
in \eqref{eq:wgcv}, while \texttt{M-sm2} 
uses a criterion similar to \eqref{eq:rcv}. 
For the local regression M-estimators of \texttt{M-loc} we used
a Tukey bisquare function and a local linear estimator. The
bandwidth was chosen using leave-one-out
cross validation with the ``ad-hoc'' robust CV criterion in
\eqref{eq:rcv2}. Finally, note that although 
the number of elements in the spline basis of \texttt{S-pen}
needs to be chosen by the user, different fits obtained with 
a wide range of basis sizes were very similar to each other. We
report here the results following the usual guideline of 
$\min( n / 4, 35) = 35$ knots placed at quantiles of
the explanatory variable (Ruppert, 2002).

\begin{figure}[htp]
\centering
\includegraphics[scale = 0.75]{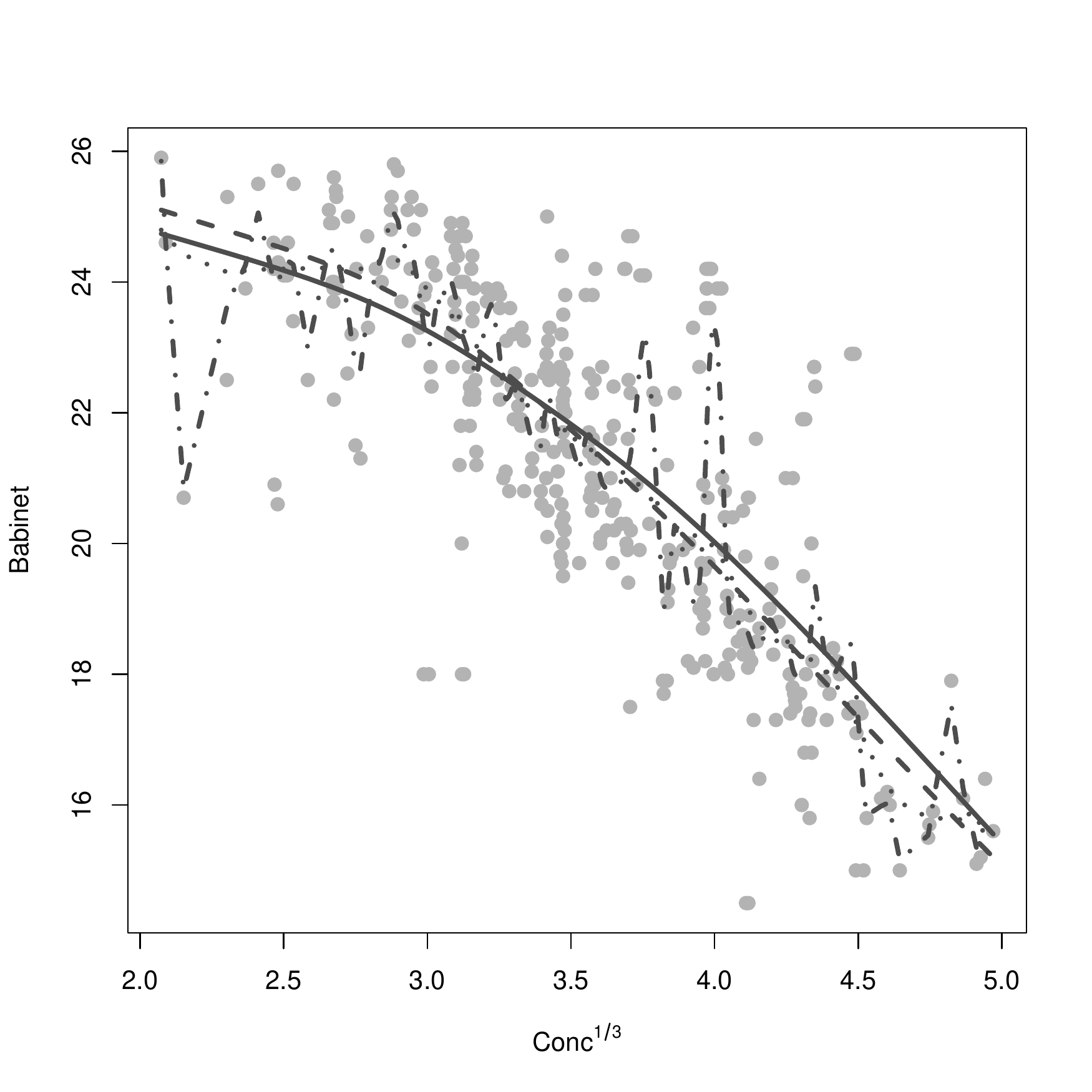}	
\caption{Robust smoothing splines fit to the Babinet data. The 
solid line is the classical fit, the dashed line is the 
\texttt{M-sm1} smoothing spline, the robust-GCV-based
M-smoothing spline \texttt{M-sm2} is the dash-dot line, 
while the dotted line corresponds to the fit obtained with
a subjectively chosen value of the smoothing parameter.}
\label{fig:1}
\end{figure}

Figure \ref{fig:1} shows the smoothing splines fits. We include
two variants of \texttt{M-sm2}. The more wiggly one was obtained
by selecting the smoothing parameter as recommended by the authors. 
The other one was obtained with a post-hoc subjective selection 
as hinted in the help page of the function \texttt{fields::qsreg}. 
Both the classical (solid line) and \texttt{M-sm1} (long 
dashes line) produce smooth fits
that fail to capture two apparent characteristics of the relationship
between these variables: they are notably lower than one would 
expect for values of $x$ around 3.0, and higher than the bulk of the
data for $x \ge 3.5$. The ``robust GCV'' based fit of \texttt{M-sm2} 
(dash-dot line) is clearly too wiggly, while the 
subjectively chosen \texttt{M-sm2} (dotted) is a 
notable improvement
(particularly near $x = 3$). Note that if one keeps forcing smoother
solutions for \texttt{M-sm2} they eventually become indistinguishable 
from \texttt{M-sm1}, as expected. 

\begin{figure}[htp]
\centering
\includegraphics[scale = 0.75]{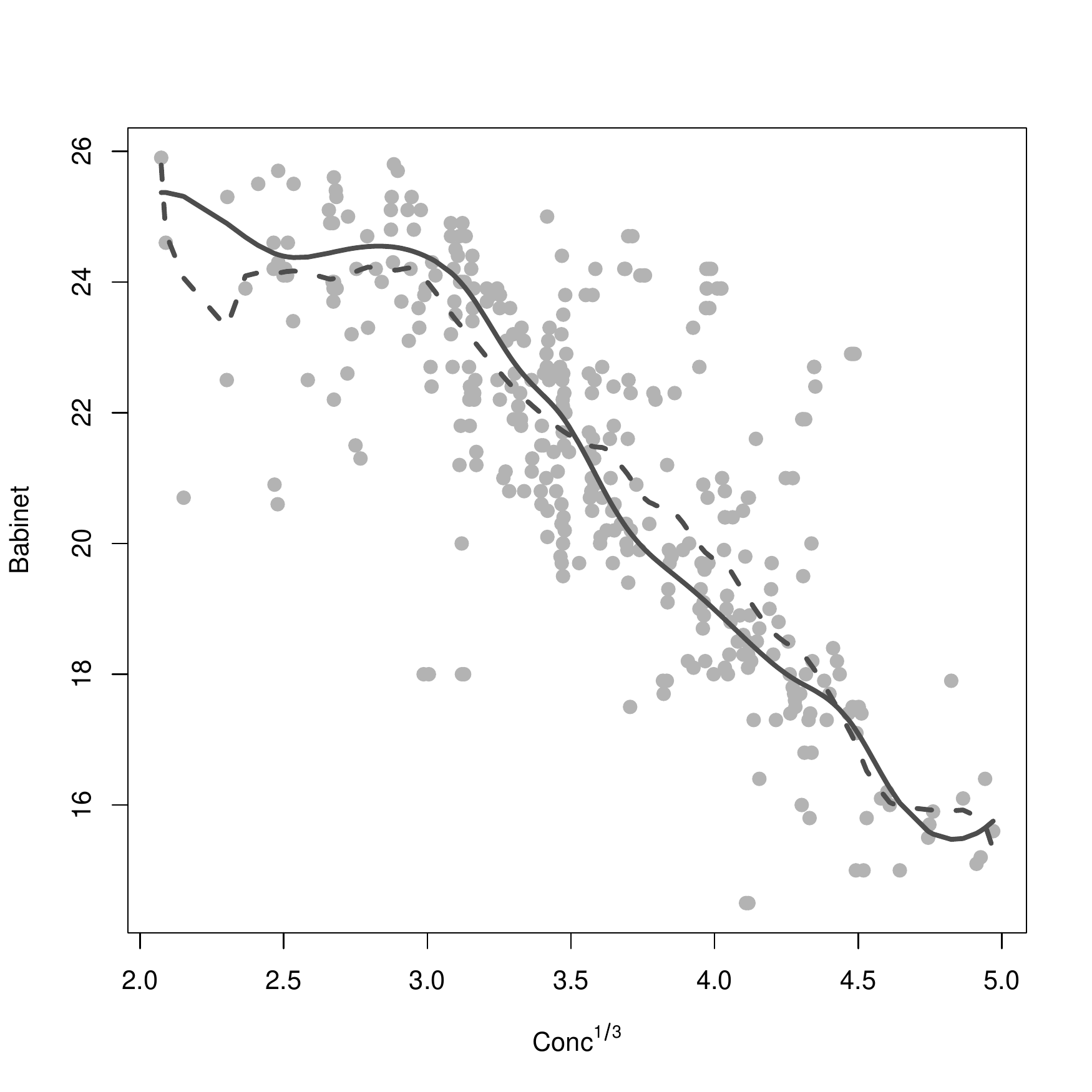}	
\caption{Robust non-parametric regression estimates for 
the Babinet data. The 
solid line is the S-penalized splines \texttt{S-pen}, 
and the dashed line is the robust local linear M-estimator
\texttt{M-loc}.}
\label{fig:2}
\end{figure}

Figure \ref{fig:2} shows the \texttt{S-pen} (solid line) and \texttt{M-loc} 
(dashed line) fits.
Note that the local linear M-estimator is less smooth than appears to 
be necessary, and also is slightly higher than expected for $x \approx 4$. 
The S-penalized spline provides both a satisfying level of smoothness
and a good fit to the trend and bulk of the data. Scripts 
reproducing this analysis can be found in an on-line supplement. 

\section{Concluding remarks}

In this review we focused on robust 
nonparametric regression estimators for models with a continuous response. 
We explicitly discussed the different practical advantages and limitations of
each approach, including options for robust bandwidth and smoothing 
parameter selection. We also reviewed recently proposed methods 
that scale well with a growing number of covariates.

Formally quantifying the level of robustness of estimation methods 
for nonparametric regression models is a challenging problem. 
Since the ``parameter'' being estimated is intrinsically 
infinite dimensional, the ``classical'' measures of robustness 
need to be extended appropriately. 
Moreover, note that 
for example, in linear regression, 
damaging outliers affect all fitted 
values, and thus, intuitively, one prefers an estimated regression 
function that ignores atypical observations 
and provides a better fit to the other
points, over an estimated regression function that may have
accommodated the outliers
and fit the other points notably worse.  
However, when the model is highly flexible, ``atypical'' observations
may not affect the quality of the fit at other
points in the training set, and thus it is hard to justify prefering
a fit that ignores outliers over one that is affected by them, particularly when 
outliers are ``isolated'' (so that there are no local ``good'' reference points).
In other words, robustness is, in fact, a concept relative to a model,
and will be affected by the ``richness'' of the model in terms 
of the regression
functions that can be estimated (the ``flexibility of the estimator'')
(see, e.g. Hable and Christmann, 2013). 
Further research on these issues would be a very welcome contribution to the
field.

\section{Acknowledgements}
The author would like to thank two anonymous referees and an Associate Editor for their
constructive comments on an earlier version of this work that resulted in a notably
improved paper. This research was supported by the Natural Sciences and Engineering Research Council of 
Canada (Discovery Grant RGPIN-2016-04288). 

\end{document}